# Temperature dependent latency of jumper cables

Florian Azendorf, Annika Dochhan, Michael Eiselt
ADVA Optical Networking SE, Meiningen, Germany, FAzendorf@advaoptical.com

## Kurzfassung

Die Temperaturabhängigkeit der Laufzeit eines optischen Signals in einem Glasfaserkabel wurde mithilfe eines Korrelations-OTDR untersucht. Während die Faser ohne Mantel einen linearen Verlauf der Laufzeit über der Temperatur aufwies, führte die Einbindung im Mantel zu einem verstärkten Ausdehnungseffekt, insbesondere im Bereich niedriger Temperaturen. Insbesondere der innere, mit der Faser verbundene Mantel (tight buffer) führt zu diesen Effekten. Bei zukünftigen, latenzsensitiven Anwendungen für 5G Netzwerke ist auf derartige Effekte Rücksicht zu nehmen.

## Abstract

The temperature dependence of the latency of an optical signal transmitted over an optical fiber was investigated by means of a correlation OTDR. While the bare fiber showed a linear latency increase over temperature, the jumper cable inside a jacket with tight buffer as inner jacket showed an increased elongation effect, especially in the range of lower temperatures. This could be attributed to the tight buffer. For future latency sensitive applications in the area of 5G networks these effects have to be taken into account.

## 1 Introduction

Synchronization in optical networks has become an important topic, as more and more time critical applications are transported in the network. With the arrival of 5G networks, the requirements in this area will become even more stringent. An application requiring a high level of synchronization accuracy is the transport of phase synchronous radio signals over a multi-core fiber or a fiber bundle to feed a phase array antenna, as envisioned in the European BlueSpace project [1]. For a carrier frequency of 26 GHz, the acceptable differential latency is limited to approximately 3 ps, corresponding to a phase error of approximately 30 degrees between two elements of the phase array. Other applications, like the transport of radio signals in digitized format using the CPRI framing, have an end-to-end latency asymmetry tolerance of 16 ns [2], leaving only a few nanoseconds or even less asymmetry budget per fiber section. This fiber latency is impacted by temperature changes of the fiber environment. A typical temperature dependence of the latency of optical fiber due to refractive index and physical length changes has been reported on the order of $7 \cdot 10^{-6}$/degC (7 ppm/degC) [3]. However, other effects, like temperature induced stress can lead to even larger temperature coefficients and lead to temperature induced latency changes on the order of several hundred picoseconds, even for a short (25 m) fiber jumper, as will be investigated in this paper.

These large dynamic changes of latency need to be monitored and compensated for in latency sensitive applications. In this paper, we will therefore also present a highly accurate correlation optical time-domain reflectometry (OTDR) technique for single-ended measurement of the fiber latency [4].

## 2 Correlation OTDR technique

The setup to characterize the fiber latency from one end is shown in **Figure 1**. A 10-Gbit/s data stream was modulated onto an optical CW wave at 1550 nm, using a Mach-Zehnder modulator. To evaluate a 25-m jumper cable, the data stream had a period of 300 ns (slightly more than the jumper cable round-trip time) and consisted of a $2^7-1$ bit PRBS sequence, appended by a sequence of zeroes.

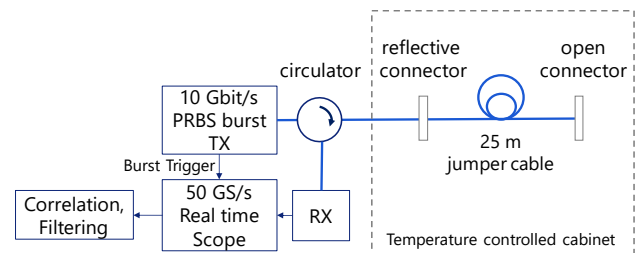

**Figure 1** Measurement setup: A 10 Gbit/s burst sequence was sent into the jumper cable with an open connector, contained in a temperature controlled enclosure. The RX consists of a 10G PIN/TIA, and the detected signal is captured by a 50-GS/s real-time oscilloscope.

The optical signal was sent into the fiber under test via a circulator. To mark the beginning of the fiber, a small air gap was provided at the input connector between circulator and jumper. The end of the fiber was left open with a non-angled connector. The reflected signal was received after the circulator and observed via a PIN/TIA receiver on a 50-GS/s real-time oscilloscope with a specified typical timing accuracy of 0.1 ppm. The oscilloscope was synchronized by the burst trigger from the transmitter, and 1000 traces were recorded and averaged by the oscilloscope. A typical averaged trace as shown in **Figure 2** contains the reflected PRBS bursts from the fiber input connector and from the fiber end. Post-processing of the averaged trace was performed in Python. The signal was correlated with the transmitted PRBS sequence and filtered to eliminate pre-and post-cursors in the correlation function due to the isolated PRBS sequence. While the sampling rate of 50 GS/s already yields a resolution of 20 ps, fitting the main correlation peaks by a Gaussian function improves the accuracy of the reflection position to a few picoseconds, as demonstrated in [4].

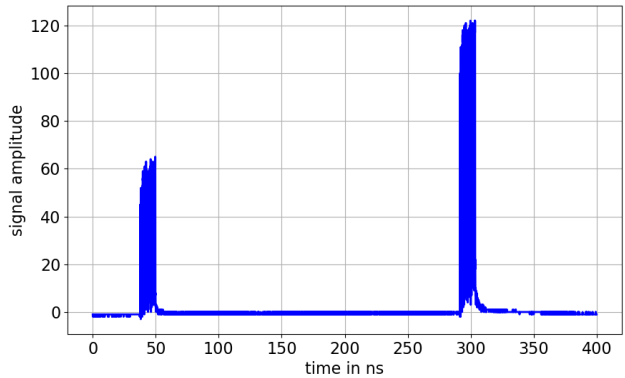

**Figure 2** Typical received reflection signal, 1000 averages.

## 3 Measurement results

The 25-m jumper under test consisted of standard single mode fiber and had an outer jacket of 2 mm diameter, with a tight buffer inner jacket. The jumper was placed in a temperature-controlled enclosure and the latency was measured over a temperature range of 10 degC to 70 degC in steps of 10 degC. The settlement time when increasing or decreasing the temperature was 20 minutes. **Figure 3** shows the transition of latency over time, when the temperature was increased from 10 degC to 20 degC.

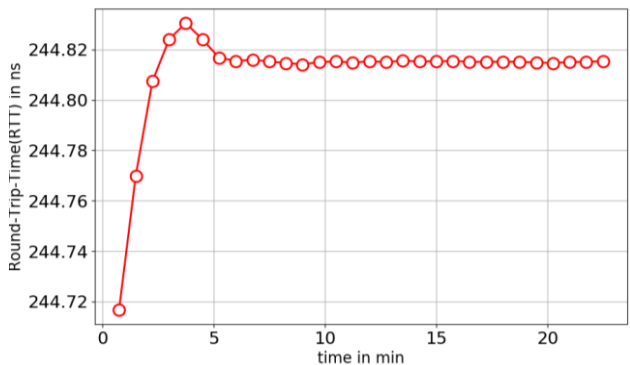

**Figure 3** Change of fiber latency over time upon temperature increase from 10 degC to 20 degC.

It can be seen that the latency increases by approximately 100 ps for a temperature increase of 10 degC and settles after about 5 minutes. The measurement was carried out for a series of increasing temperatures and immediately afterwards for decreasing temperatures from 70 degC to 10 degC in steps of 10 degC, respectively. **Figure 4** shows the evolution of the round-trip latency over temperature. It can be seen that the temperature coefficient of the cable was not constant but decreased with increasing temperature. For low temperatures, the temperature coefficient was measured to be approximately 40 ppm/degC, which is approximately six times the expected value. Over the full temperature range, the average temperature coefficient was approximately 17 ppm/degC. The measured latency was also dependent on the direction of the temperature change with a characteristics inverse to a hysteresis effect: When reducing the temperature, the latency was decreasing already for higher temperature values than for the increasing temperature series.

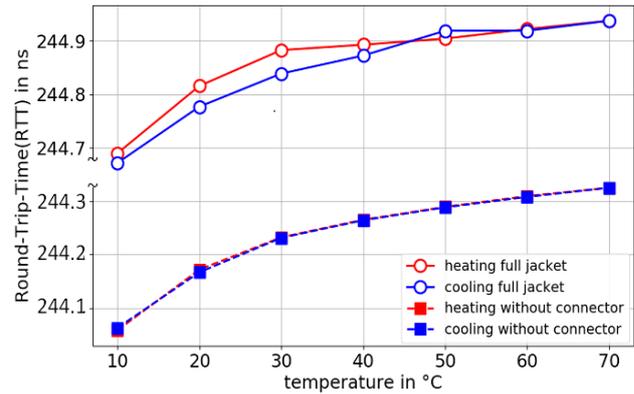

**Figure 4** Temperature dependence of latency of 25-m jumper cable with outer jacket and connectors (solid lines) and with one connector cleaved off (dashed lines), for increasing (red) and decreasing (blue) temperatures.

To investigate, how the outer jacket contributed to the fiber latency change, e.g. by introducing tensile stress, we removed the open connector at the cable end by a straight cleave. Cutting off the connector at the fiber end allowed the fiber to move in the outer jacket, releasing some of the tensile stress. Still, the fiber was placed in an inner jacket (tight buffer type), where friction might lead to tensile stress. The measurement series were repeated with increasing and decreasing temperatures, resulting in latencies as shown in the second set of traces in Figure 4. Due to cutting off 6 cm of the fiber, the round-trip latency was reduced by approximately 600 ps. As can be seen, the general, non-linear dependence of latency over temperature was still maintained with an average temperature coefficient over the 60-degC temperature range of 17.8 ppm/degC. However, the flat region around 50 degC was smoothed out, and the inverse hysteresis effect was not observed anymore.

The results obtained with a jacketed fiber were compared to a bare fiber without jacket. **Figure 5** shows the temperature dependence of the latency of an equivalent 25-m piece of an un-spooled standard single mode fiber. The curve exhibits a linear behavior with a latency variation of 105 ps over 60 degC temperature variation, i.e. a temperature coefficient of 6.9 ppm/degC, as expected from previous publications [3]. This points to the strong impact of the jacket to the latency behavior of optical transmission fiber.

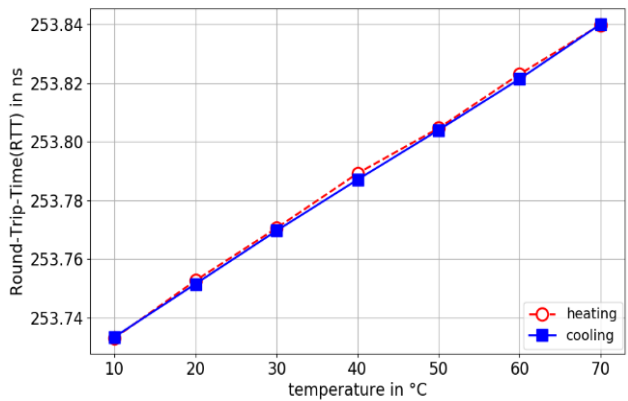

**Figure 5** Temperature dependence of latency of 25-m bare fiber for increasing (red) and decreasing (blue) temperatures.

## 4 Summary

We investigated the temperature dependence of a jumper cable with outer buffer by using a single ended correlation OTDR technique. The complete 25-m jumper exhibited a 266-ps latency variation over a temperature range of 60 degC with a temperature coefficient of 40 ppm/degC for low temperatures, decreasing with increasing temperature. Also, an inverse hysteresis behavior was observed. This was partly attributed to tensile stress caused by the outer jacket expansion. After removing the fiber end connector and therefore the connection to the outer buffer, the inverse hysteresis was eliminated, but the decreasing temperature coefficient was still observed. This might be attributed to tensile stress, still present due to friction of the fiber in the inner jacket. Comparing the results to an unjacketed fiber, the latter showed a more linear latency vs. temperature evolution with a temperature coefficient of 6.9 ppm/degC. The strong impact of the jacket on the latency behavior of optical transmission fiber should be taken into account for cabled fiber, when latency variations are considered for latency critical variations.

## 5 Acknowledgement

This project has received funding from the European Union's Horizon 2020 research and innovation programme under grant agreement No 762055.